\documentclass[nofootinbib,onecolumn,preprint,showpacs,superscriptaddress]{revtex4-1}

\usepackage[utf8]{inputenc}
\usepackage{mathtools}
\usepackage{graphicx}
\usepackage{caption}
\usepackage{subcaption}
\usepackage{xcolor}
\usepackage{soul}
\usepackage{enumerate}
\usepackage{amsmath,amsfonts,amssymb,amsthm}
\usepackage{commath}
\usepackage{indentfirst}
\usepackage{setspace}
\usepackage{dsfont}
\usepackage{lastpage}
\usepackage{float}
\usepackage{epstopdf}
\usepackage{scrextend}
\usepackage{hyperref}
\usepackage{footnote}
\usepackage[ruled,vlined]{algorithm2e}
\usepackage{times}
\usepackage{feynmf}
\usepackage{physics}
\usepackage{simpler-wick}
\usepackage{cancel}
\usepackage{comment}
\usepackage{subcaption}
\usepackage{booktabs}
\usepackage{mathtools}
\usepackage{natbib}
\usepackage{orcidlink}

\begin{document}

\def\a{\alpha}
\def\b{\beta}
\def\G{\Gamma}
\def\t{\theta}
\def\s{\sigma}
\def\d{\partial}
\def\ddl{\mathcal{L}}
\def\ddlint{\ddl_{\text{int}}}
\def\ddlem{\frac{1}{4}F_{\mu\nu}F^{\mu\nu}}
\def\l{\left}
\def\r{\right}
\def\L{\Lambda}
\def\meio{\frac{1}{2}}
\def\ceb{\coth\left(n \pi \frac{E}{B} \right)}
\def\cbe{\coth\left(n \pi \frac{B}{E} \right)}
\def\mg{\mathcal{G}}
\def\mf{\mathcal{F}}
\def\Msun{M_{\odot}}
\def\Qlim{Q_{\mathrm{lim}}}
\def\ra{\rightarrow}
\def\o{\mathcal{O}}
\def\aring{\stackrel{\circ}}
\def\rmunu{R_{\mu\nu}}\textbf{}
\def\espacinho{\vspace{0.4cm}}
\def\jumph{\hspace{0.2cm}}
\def\jumpheq{\hspace{0.2cm}=\hspace{0.2cm}}
\def\sen{\sin}
\def\checkmark{\tikz\fill[scale=0.4](0,.35) -- (.25,0) -- (1,.7) -- (.25,.15) -- cycle;} 
\def\vp{\vec{p}}
\def\vk{\vec{k}}
\def\vd{\vec{d}}
\def\vx{\vec{x}}
\def\vy{\vec{y}}
\def\vq{\vec{q}}
\def\vr{\vec{r}}
\def\dtq{\frac{d^3q}{(2\pi)^3}}
\def\dtfourk{\frac{d^4k}{(2\pi)^4}}
\def\empty{\hspace{0.15cm}}
\def\psibar{\Bar{\psi}}
\makeatletter
\newcommand{\vast}{\bBigg@{3}}
\newcommand{\Vast}{\bBigg@{4}}
\newcommand{\Vaster}{\bBigg@{4}}
\let\oldref\ref
\renewcommand{\ref}[1]{(\oldref{#1})}

\title{Spacetime curvature corrections for the Yukawa potential and its application for the Reissner-Nordström Metric} 

\author{J. V. Zamperlini\orcidlink{0009-0002-9702-1555}}
\email{joao.zamperlini@posgrad.ufsc.br}
\affiliation{Departamento de Física, CFM - Universidade Federal de \\ Santa Catarina; C.P. 476, CEP 88.040-900, Florianópolis, SC, Brazil}

\author{C. C. Barros Jr. \orcidlink{https://orcid.org/0000-0003-2662-1844}}
\email{barros.celso@ufsc.br}

\affiliation{Departamento de Física, CFM - Universidade Federal de \\ Santa Catarina; C.P. 476, CEP 88.040-900, Florianópolis, SC, Brazil} 

\begin{abstract}
In this paper, we investigate the influence of the spacetime curvature on the Yukawa potential, focusing on boson-boson interactions derived from the \(\phi^3\) theory. Using the Bunch-Parker propagator expansion within Born's first approximation, we derive a Yukawa-like potential in a curved spacetime. We analyze the impact of the curvature on the propagator in momentum space, revealing modifications to the potential and showing that the corrections are determined by geometric quantities from Einstein's equations, like the Ricci scalar and tensor. We illustrate this using the Reissner-Nordström metric, highlighting the corrections' magnitude for specific parameters. Our results underscore the nuanced interplay between spacetime curvature and quantum interactions, providing insights into nucleon-nucleon systems in curved spacetimes or near strong gravitational fields.
\end{abstract}

\maketitle

\section{Introduction}

The influence of the spacetime curvature, described by Einstein's Theory of General Relativity, on particles and fields remains a subject of study and research. A question of interest is how strong gravitational fields influence the interactions described by quantum fields and if this kind of influence is significant in these physical processes. Developments about this question bring light to the most elementary formulation of the physical theory. 

In recent years, a large amount of studies considering quantum systems in different spacetimes have been performed with this purpose. The Klein-Gordon and Dirac equations in curved spacetimes  \cite{Parker:1980hlc}  have been investigated in several types of backgrounds, as for example in the Schwarzschild metric \cite{elizalde_1987}, Kerr black holes \cite{chandra}, in the Hartle-Thorne spacetime \cite{Pinho:2023nfw} and considering cosmic strings \cite{Santos:2016omw}, \cite{Santos:2017eef}, \cite{Vitoria:2018its}. Quantum oscillators \cite{Ahmed:2022tca}, \cite{Ahmed:2023blw}, \cite{Santos:2019izx}, \cite{Yang:2021zxo}, \cite{Soares:2021uep}, \cite{Rouabhia:2023tcl},
the Casimir \cite{Santos:2018jba} and other effects \cite{Sedaghatnia:2019xqb}, \cite{Guvendi:2022uvz},
\cite{Vitoria:2018mun}, \cite{Barros:2004ta} have also motivated many works. These papers show many aspects and results of the formulation of quantum systems in the context of general relativity.

In 1979, in the interesting paper presented by Bunch and Parker \cite{BunchParkerProp}, the Feynman propagator in the momentum space has been calculated in a curved spacetime. The Riemann normal coordinates have been employed in order to allow the utilization of the Minkowski space techniques to deal with the divergences in curved spaces.  
For this reason, this seminal work served as a reference for several articles featuring a reliable formulation of Green's functions. In particular, more recently, through this result, it was possible to obtain an expression for the S-matrix in a curved spacetime \cite{SMatrixCurvedMandal_2021} and in \cite{Kogler:2022byd}
a study was carried out on the quantum consistency in the formation of black holes. Obtaining potentials corrected by the curvature of spacetime, or by General Relativity, has already been developed in the literature. Using these ideas in order to obtain an effective field theory, the potential through the exchange of gravitons was obtained up to 1-loop 
\cite{GravitonNewtonMassesBjerrum-Bohr:2002gqz}
and in \cite{BornApproximationDeSitterFerrero:2021lhd} with the Born approximation in
the de Sitter space and these results show the corrections to the
Newtonian potential up to the first order. However, there are no studies addressing how other potentials of other interactions are affected by a non-Minkowskian metric.

So, adopting this framework, it is possible formulate interactions of particles by considering the quantization of the fields. In this work we will construct a Yukawa-like potential based on a $\phi^3$ theory, that may be used to describe an
one-pion exchange interaction, with the corrections to the propagator in a curved space given by the formalism presented in \cite{BunchParkerProp}. 
The study of the Yukawa potential within the framework of general relativity holds significant importance in contemporary theoretical physics. This potential, foundational to our understanding of nuclear interactions, delineates the short-range behavior of fundamental forces within atomic nuclei. For simplicity, for this study, the interaction of pions considering the $\phi^3$ vertex will be taken into account, and it is a reasonable way to manage this subject as it will provide all the important aspects of this interaction, and the generalization to the nucleon-nucleon system is straightforward. In fact, with this procedure, we can obtain the same analytical form that is derived for the central component of the nucleon-nucleon potential, taking into account one-pion exchange. For this reason, considering the correct parameters, it is possible to estimate, at least in a first approximation, the corrections for the nuclear Yukawa potential due to the structure of the spacetime.   As a choice, we will investigate the effect of the spacetime of a Reissner-Nordström black hole
by calculating the corrections that appear in the potential and then searching for 
the values of the parameters that may provide a significant effect.

The structure of this paper is as follows: In Sec. II we will show how the propagator in a curved spacetime may be calculated and in Sec. III a revision of the Yukawa potential in the Minkowski space will be made and then it will be calculated in an arbitrary spacetime. In Sec. IV
the corrections for the Reissner-Nordström metric will be obtained, in Sec. V the results will be presented and in Sec. VI the conclusions will be drawn.

\section{ Feynman propagator in curved spacetime}

In a general way,
under the formalism of quantum fields in curved spaces the influence of a non-trivial metric on Feynman diagrams is related to the alteration of the propagator, as expected, since it describes the propagation amplitude between the spacetime points $x_1$ and $x_2$ and then it must be sensible to the spacetime geometry. In \cite{BunchParkerProp, BunchProp2RenormPhi4:1981tr} the two point Green function $G_F(x_1,x_2)$ (directly related with the Feynman propagator by a factor $-i$) in a curved spacetime is investigated and for a scalar particle is given by \cite{book:BirrellDavies}
\begin{equation}
    \l(\Box_c + m^2 \r)G_F(x,x') = -\frac{\delta^{(4)}\l(x-x'\r)}{\sqrt{-g}} \quad,
\end{equation}
with $\Box_c=g^{\mu\nu}\d_\mu\d_\nu$ being the curved space d'Alambertian and $m$ is the $\phi$ boson mass. The procedure to be followed involves in taking an infinite expansion for $\Box_c$ using the identities from Riemann normal coordinates and also for the propagator in momentum space, that is given by
\begin{equation}
    G(y) = \int \dtfourk e^{i k \cdot y}G(k) = \int \dtfourk e^{i k \cdot y}\l[G_0(k) + G_1(k) + G_2(k) + \cdots \r] \quad,
\end{equation}
where $G_0(k)$ is the flat-space propagator and $G_j(k)$ for $j = 1,2,...$ represent curved space corrections in momentum space, that can be found in an iterative way. Keeping up to first order in the Ricci tensor, the expression for the momentum space Green's function $G(k)$ under the Wick rotation procedure can be given by \cite{book:ShapiroGravity}:
\begin{equation}
    G(k) = \frac{1}{k^2+m^2} + \frac{1}{3}\frac{R'}{\l(k^2+m^2\r)^2} - \frac{2}{3}\frac{R'^{\mu\nu}k_\mu k_\nu}{\l(k^2+m^2\r)^3} + \mathcal{O}\l(k^{-5}\r) \qquad,
    \label{eq:propEspCurvoMom}
\end{equation}
where $R'^{\mu\nu}$ and $R'$ are respectively the Ricci tensor components and the Ricci scalar calculated at the spacetime point $x'$ where the expansion is evaluated.

From this result we may determine the expansion of the Green's function
in terms of the geometric quantities for some generic metric $g_{\mu\nu}$ associated with curved space corrections to quantum potentials, such as the Yukawa potential originated from the $\phi^3$ theory. In the following sections the implications of these results will be studied.

\section{$\phi^3$ Yukawa Potential in a Curved Spacetime from Born's First Approximation}

\begin{figure}[!h]
    \centering
    \includegraphics[scale=0.3]{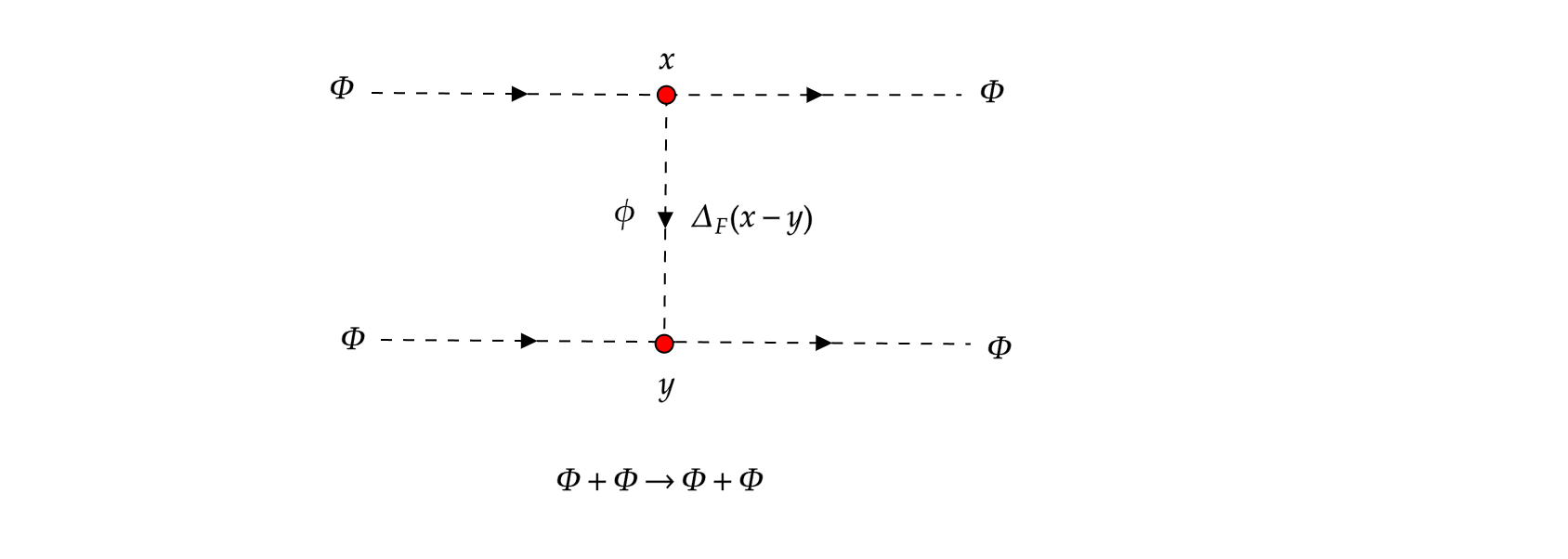}
    \caption{Leading order Feynman diagram for the process $\Phi+\Phi \ra \Phi+\Phi$.}
    \label{fig:diagramaphi3LO}
\end{figure}

First introduced in 1935 \cite{Yukawa:1935xg}, the Yukawa potential represents a fundamental contribution to our understanding of nuclear forces, postulating a mediating particle, the meson, responsible for conveying the strong nature of these interactions. So it is interesting to investigate how this kind of interaction may be affected by the metric. For this purpose, in this section we will consider a model for the interaction of mesons based on the $\phi^3$ theory \cite{book:Gross} by taking into account the lagrangian
\begin{equation}
    \mathcal{L} = \frac{1}{2} (\partial_{\mu}\phi)^2 - \frac{\mu^2}{2} \phi^2 + \frac{1}{2} \partial_{\mu}\Phi^*\partial^{\mu}\Phi - \frac{m_{\Phi}^2}{2} |\Phi|^2 - \frac{\lambda}{3!} \Phi^{*}\phi\Phi \qquad ,
    \label{eq:lagrangianphi3}
\end{equation}
and supposing the interaction at the tree level as it is shown in  the diagram of  \autoref{fig:diagramaphi3LO}.

\noindent
With this procedure it is possible to obtain a Yukawa-like potential. In this work, for simplicity, we will only take into account the interactions of mesons that is a
straightforward approach to identify the essential features
of this kind of interaction and allows us to calculate the corrections originated by the metric. The nucleon-nucleon interaction is left for future works.

From the Feynman diagram for the scattering 
$\Phi+\Phi \ra \Phi+\Phi$
depicted in \autoref{fig:diagramaphi3LO} 
the amplitude is given by
\begin{equation}
    \mathcal{M} = i \l(-i\lambda\r)^2\l(\frac{-i}{\mu^2-q^2}\r) \qquad ,
    \label{eq:amplitudephi3LO}
\end{equation}

\noindent
which in the Born approximation is related by the potential
in the 3-momentum space, in a non-relativistic approximation,
by the expression \footnote{Considering $E_{\Phi} \approx m_{\Phi}.$}:
\begin{equation}
    \Tilde{V}\l(|\vq|\r) = \frac{\mathcal{M}}{4m_{\Phi}^2} \qquad.
    \label{eq:BornVq}
\end{equation}

Performing the inverse Fourier transform on eq.\ref{eq:BornVq} we obtain the Yukawa potential in a flat space
\begin{equation}
    V(r) = \int \dtq e^{i\vq\cdot\vr} \frac{\mathcal{M}}{4m_{\Phi}^2} = 
    -\frac{\lambda^2}{4m_{\Phi}^2}\int \dtq e^{i\vq\cdot\vr} \frac{1}{\mu^2+|\vq|^2} = 
    -\l(\frac{\lambda^2}{4\pi}\r)\frac{1}{4m_\Phi^2}\frac{e^{-\mu r}}{r} \quad,
    \label{yukawaphi3potential}
\end{equation}
which, as it is well known, show the attractive behavior of the interaction whose effective range depends upon the scale of the mediating boson mass $\mu$. We must notice, that despite the fact that this potential has been obtained in the 
$\phi^3$ theory, it presents the same analytical form of the central potential of the nucleon-nucleon interaction. So with the correct parameters the results may be used to estimate the corrections for this potential.

Now, for a interaction in a curved spacetime, in the Riemann
normal coordinates, according to eq. \ref{eq:propEspCurvoMom} we have 
\begin{equation}
    V(\vr) = -\frac{\lambda^2}{4m_\Phi^2}\int \dtq e^{i\vq\cdot\vr}\l[\frac{1}{\mu^2+|\vq|^2} + \frac{1}{3}\frac{R'}{\l(\mu^2+|\vq|^2\r)^2} - \frac{2}{3}\frac{R'_{\mu\nu}q^\mu q^\nu}{\l(\mu^2+|\vq|^2\r)^3} +\cdots \r] \quad.
    \label{PotentialCurvedSpaceInts}
\end{equation}
We bring our attention that in this work we will just keep the expansion of the potential given by eq.\ref{PotentialCurvedSpaceInts} up to first order in the Ricci tensor components as far as the corrections are very small. Higher-order terms would lead to corrections given by higher powers of $G$, resulting in even smaller corrections, the numerical results found in Sec. V justify this assumption.

The first term in the brackets in \autoref{PotentialCurvedSpaceInts} gives us the flat-space Yukawa potential $V_0(r)$
\begin{equation}
    V_0(r) = -\frac{\lambda^2}{4m_\Phi^2}\int \dtq e^{i\vq\cdot\vr}\l[\frac{1}{\mu^2+|\vq|^2}\r]  \ ,
\end{equation}
\noindent
 that after the integration is
 \begin{equation}
     V_0(r) = -\l(\frac{\lambda^2}{4m_\Phi^2} \frac{1}{4\pi}\r)\frac{e^{-\mu r}}{r}  \quad  ,
     \label{V0}
 \end{equation}
\noindent
whereas the second term weighted by the Ricci scalar at the point $x'$ is simply the integral of the propagator squared, 
that gives
\begin{equation}
    -\frac{\lambda^2}{4m_\Phi^2}\int \dtq e^{i\vq\cdot\vr}\l[\frac{1}{3}\frac{R'}{\l(k^2+m^2\r)^2}\r] = -\l(\frac{\lambda^2}{4m_\Phi^2} \frac{1}{4\pi} \r)\frac{e^{-\mu r}}{r} \cdot \l[\frac{1}{3}R' \r] \cdot\frac{r}{2\mu} \quad.
\end{equation}

As a matter of fact, all the next corrections in \ref{PotentialCurvedSpaceInts}, which are integrals of powers of the flat-space propagator, will result in the flat-space Yukawa potential multiplied by a function of $\mu$ and $r$, in other words, we can always reduce such integrals as
\begin{equation}
    \int \dtq \frac{e^{i\vq\cdot\vr}}{\l(\mu^2 + |\vq|^2\r)^n} = \frac{1}{(n-1)!} \l(-\frac{1}{2\mu}\frac{\d}{\d \mu}\r)^{n-1} \l(\frac{1}{4\pi}\frac{e^{-\mu r}}{r} \r) \quad.
\end{equation}

The third term in eq. \ref{PotentialCurvedSpaceInts},  which is the second correction, is proportional to the Ricci tensor components and to the respective momenta components. This integral is made simple if expressed in Cartesian coordinates
\begin{align*}
    I_3 &= \int \dtq e^{i\vq\cdot\vr}\frac{R'_{\mu\nu}q^\mu q^\nu}{\l(\mu^2+|\vq|^2\r)^3} \\ 
    &= \int \dtq \frac{e^{i(q^xx+q^yy+q^zz)}}{\l(\mu^2+|\vq|^2\r)^3} \biggr[R'_{xx}q^xq^x + R'_{yy}q^yq^y + R'_{zz}q^zq^z +2R'_{xy}q^xq^y + 2R'_{xz}q^xq^z + 2R'_{yz}q^yq^z\biggr] \quad,
\end{align*}
and after integration is 
\begin{equation}
    \begin{aligned}
    I_3 = \l(\frac{1}{4\pi}\frac{e^{-\mu r}}{r}\r)\biggr[& \l( R'_{xx} + R'_{yy} + R'_{zz} \r)\mu r -R'_{xx} \mu^2x^2 - R'_{yy}\mu^2y^2 - R'_{zz}\mu^2z^2 + \\ & -2R'_{xy}\mu^2xy - 2R'_{xz}\mu^2xz - 2R'_{yz}\mu^2yz\biggr] \frac{1}{8\mu^2} \quad.
\end{aligned}
\end{equation}

Substituting it back in eq. \ref{PotentialCurvedSpaceInts} we obtain an expression in terms of the flat-space potential $V_0(r)$ 
\begin{equation}
\begin{aligned}
    V(\vr) = V_0(r) \vast\{ & 1 + \l(\frac{1}{3} R' \r)\frac{\mu r}{2\mu^2} + \frac{1}{3}\biggr[\l(R'_{xx} + R'_{yy} + R'_{zz}\r)\mu r + \\ & -R'_{xx} \mu^2x^2 - R'_{yy}\mu^2y^2 - R'_{zz}\mu^2z^2 +\\ & -2R'_{xy}\mu^2xy - 2R'_{xz}\mu^2xz - 2R'_{yz}\mu^2yz\biggr] \frac{1}{4\mu^2} \vast\} \quad.
    \label{cartesianCurvedSpacePotentialResult}
\end{aligned}
\end{equation}

The second correction term shows an interesting feature, the integral in the cubic power of the meson propagator is weighted by the Ricci tensor components multiplied by the momenta components in the respective directions, resulting in the explicit radial symmetry breaking of the Yukawa potential. That means, if the spacetime is equipped with a metric tensor in which the curvature tensor components are different for the three Cartesian spatial directions, there is a break in radial symmetry as a consequence, resulting in a Cartesian-coordinate-dependent potential energy which can vary based on the chosen metric.

Another observation to be taken is related to bound states, since $V_0(r)$ is attractive \ref{yukawaphi3potential}, it is possible that this  kind  of states exists. Since the curved space correction terms in \ref{cartesianCurvedSpacePotentialResult} can end up having a positive or negative sign, that could be understood as a form of strengthening or weakening of the interaction respectively, which should be analyzed for the particular metric on which the result would be applied to. 

Observing the resulting expression, it is possible that these corrections become important if the quantum system is near a body which is source of an intense gravitational field, such as in the early universe or in the vicinity of black holes and neutron stars. As a matter of better understanding this expression, we will apply the obtained results for the Reissner-Nordström metric in the next section.

\section{$V(\vr)$ in the Reissner-Nordström Metric}

The results obtained in the previous section show the lowest order corrections for the potential. Now, in order to obtain numerical results,
we will study one of the simplest metrics that presents 
 a non-null Ricci tensor, and therefore brings correction to the Yukawa potential up to the order of the expansion we have truncated, that
 is the charged black hole metric or Reissner-Nördstorm (RN) metric \cite{Reissner1916}. This metric is given by the line element
\begin{equation}
    \mathrm{d}s^2 = -\left(1-\frac{r_s}{r}+\frac{r_Q^2}{r^2}\right)\mathrm{d}t^2 + \l(1-\frac{r_s}{r}+\frac{r_Q^2}{r^2}\r)^{-1}\mathrm{d}r^2 + r^2\mathrm{d}\theta^2 + r^2 \sin^2(\theta) \mathrm{d}\varphi^2 \quad,
    \label{eq:RNmetriclineelement}
\end{equation}
on which
\begin{equation}
    r_Q^2=\frac{Q^2 G}{4 \pi} \qquad \text{and} \qquad r_s=2 G M \quad ,
\end{equation}
where $Q$ is the electric charge and $M$ is the black hole mass.
From this section onward we use the natural units system:
\begin{equation}
    \begin{aligned}
        c=\hbar=1 \qquad \text{and} \qquad G=m_P{}^{-2}= 6.7087\times 10^{-39} \text{GeV}^{-2} \quad ,
    \end{aligned}
\end{equation}
where $m_P$ is the Planck mass.

The Ricci tensor for the RN metric can be calculated from
eq. \ref{eq:RNmetriclineelement} and written in matrix form
\begin{equation}
R_{\mu \nu}=\l(\begin{array}{cccc}
\frac{r_Q^2}{r^4}-\frac{r_Q^2 r_s}{r^5}+\frac{r_Q^4}{r^6} & 0 & 0 & 0 \\
0 & \frac{r_Q^2}{r^3 r_s-r^4-r^2 r_Q^2} & 0 & 0 \\
0 & 0 & \frac{r_Q^2}{r^2} & 0 \\
0 & 0 & 0 & \frac{r_Q^2}{r^2} \sen ^2 \theta
\end{array}\r) \quad,
\end{equation}
and the corresponding Ricci scalar $R=g^{\mu\nu}R_{\mu\nu}$ is zero.

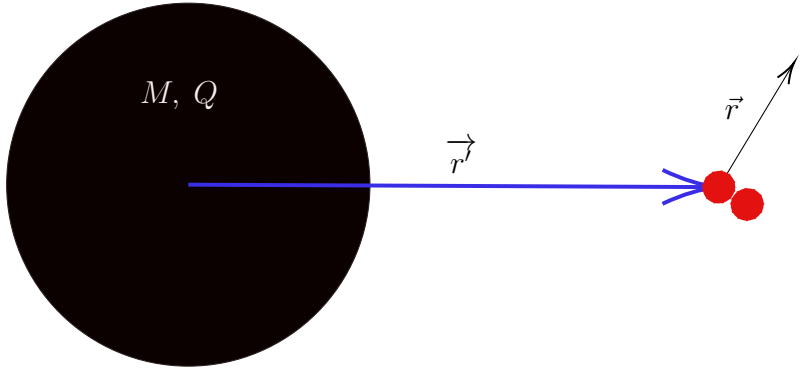
\begin{figure}[!h]
    \centering
    \begin{tikzpicture}[x=0.75pt,y=0.75pt,yscale=-1,xscale=1]
    
    \draw  [color={rgb, 255:red, 42; green, 33; blue, 33 }  ,draw opacity=1 ][fill={rgb, 255:red, 12; green, 1; blue, 1 }  ,fill opacity=1 ] (44,138.61) .. controls (44,88.01) and (85.01,47) .. (135.61,47) .. controls (186.2,47) and (227.22,88.01) .. (227.22,138.61) .. controls (227.22,189.2) and (186.2,230.22) .. (135.61,230.22) .. controls (85.01,230.22) and (44,189.2) .. (44,138.61) -- cycle ;
    \draw [color={rgb, 255:red, 58; green, 45; blue, 230 }  ,draw opacity=1 ][fill={rgb, 255:red, 139; green, 100; blue, 100 }  ,fill opacity=1 ][line width=1.5]    (135.61,138.61) -- (400.22,139.85) ;
    \draw [shift={(403.22,139.86)}, rotate = 180.27] [color={rgb, 255:red, 58; green, 45; blue, 230 }  ,draw opacity=1 ][line width=1.5]    (28.42,-8.55) .. controls (18.07,-3.63) and (8.6,-0.78) .. (0,0) .. controls (8.6,0.78) and (18.07,3.63) .. (28.42,8.55)   ;
    \draw    (403.22,139.86) -- (440.18,78.58) ;
    \draw [shift={(441.22,76.86)}, rotate = 121.1] [color={rgb, 255:red, 0; green, 0; blue, 0 }  ][line width=0.75]    (10.93,-3.29) .. controls (6.95,-1.4) and (3.31,-0.3) .. (0,0) .. controls (3.31,0.3) and (6.95,1.4) .. (10.93,3.29)   ;
    \draw  [color={rgb, 255:red, 239; green, 11; blue, 11 }  ,draw opacity=1 ][fill={rgb, 255:red, 228; green, 17; blue, 17 }  ,fill opacity=1 ][dash pattern={on 0.84pt off 2.51pt}] (395.01,139.64) .. controls (395.27,135.11) and (399.17,131.54) .. (403.7,131.66) .. controls (408.24,131.79) and (411.69,135.56) .. (411.43,140.09) .. controls (411.16,144.62) and (407.26,148.19) .. (402.73,148.07) .. controls (398.2,147.94) and (394.74,144.17) .. (395.01,139.64) -- cycle ;
    \draw  [color={rgb, 255:red, 239; green, 11; blue, 11 }  ,draw opacity=1 ][fill={rgb, 255:red, 228; green, 17; blue, 17 }  ,fill opacity=1 ][dash pattern={on 0.84pt off 2.51pt}] (409.67,146.15) .. controls (411.17,141.87) and (415.89,139.5) .. (420.22,140.86) .. controls (424.54,142.23) and (426.83,146.8) .. (425.34,151.09) .. controls (423.84,155.37) and (419.11,157.74) .. (414.79,156.37) .. controls (410.46,155.01) and (408.17,150.43) .. (409.67,146.15) -- cycle ;
    
    \draw (405,93) node [anchor=north west][inner sep=0.75pt]    {$\vec{r}$};
    \draw (264,114) node [anchor=north west][inner sep=0.75pt]    {$\overrightarrow{r'}$};
    \draw (110,85) node [anchor=north west][inner sep=0.75pt]  [color={rgb, 255:red, 245; green, 237; blue, 237 }  ,opacity=1 ]  {$M,\ Q$};
    
    \end{tikzpicture}
    \caption{Pictorial representation of the position vectors of the two scalar particles system relative to the charged black hole.}
    \label{fig:buraconegroNN}
\end{figure}

For the calculations we will assume the coordinates shown in \autoref{fig:buraconegroNN}. The origin of the potential will be located at the position $\vr'$ from the center of the black hole and the point where the potential is calculated is represented by the variable 
$\vr$. So, eq. \ref{cartesianCurvedSpacePotentialResult} 
will be used to calculate $V(\vr)$ with the Ricci tensor and scalar evaluated at $\vr'$.

\tikzset{every picture/.style={line width=0.75pt}} 

To evaluate eq. \ref{cartesianCurvedSpacePotentialResult} we need to transform the Ricci components from Cartesian to spherical coordinates. This transformation may be done accordingly with 
\begin{equation}
    R_{\a\b}= \frac{\d x^\mu}{\d x^\a}\frac{\d x^\nu}{\d x^\b} R_{\mu\nu}  
\end{equation}
and taking into account the coordinates shown in \autoref{fig:buraconegroNN}. In this case, for a general metric with a diagonal Ricci tensor, the potential is given by
{\footnotesize{
\begin{equation}
\begin{aligned}
V(\vec{r}) = V_0(r) \vast\{& 1  + \frac{1}{12}\Biggr[\l(2 R' +R'_{rr} + \frac{1}{r'^2}R'_{\theta\theta} + \frac{1}{r'^2\sen^2(\theta')}R'_{\varphi\varphi}\r)\frac{ r}{\mu} + \\& -\l(\sen^2(\theta')\cos^2(\varphi') R'_{rr} + \cos^2(\theta')\cos^2(\varphi')\frac{R'_{\theta\theta}}{r'^2}  + \sen^2(\varphi')\frac{R'_{\varphi\varphi}}{r'^2\sen^2(\theta')} \r) x^2 + \\& - \l(\sen^2(\theta')\sen^2(\varphi') R'_{rr} + \cos^2(\theta')\sen^2(\varphi')\frac{R'_{\theta\theta}}{r'^2} + \cos^2(\varphi')\frac{R'_{\varphi\varphi}}{r'^2\sen^2(\theta')} \r)y^2 + \\& - \l(\cos^2(\theta') R'_{rr} + \sen^2(\theta')\frac{R'_{\theta\theta}}{r'^2}\r)z^2 + \\& - \sen(2\varphi')\l( \sen^2(\theta') R'_{rr} + \cos^2(\theta')\frac{R'_{\theta\theta}}{r'^2}  - \frac{R'_{\varphi\varphi}}{r'^2\sen^2(\theta')} \r)xy + \\& - \sen(2\theta')\l(  R'_{rr} - \frac{R'_{\theta\theta}}{r'^2} \r)\l(\cos(\varphi')x+\sen(\varphi')y\r)z \Biggr]\vast\} \qquad. \label{eq:potencialgeralRicciEsfericas}
\end{aligned}    
\end{equation}
}}
which for the particular case of the RN metric is
\footnotesize{\begin{equation}
    \begin{aligned}
    V(\vec{r}) = V_0(r) \vast\{& 1 +
    \frac{1}{12}\Biggr[
    \frac{r_Q^2}{r'^3 r_s-r'^4-r'^2 r_Q^2}\Biggr(\frac{\mu r}{\mu^2} - \biggr( \sen^2(\theta')\cos^2(\varphi') x^2 + \sen^2(\theta')\sen^2(\varphi')y^2 + \\& + \cos^2(\theta')z^2 + \sen^2(\theta')\sen(2\varphi') xy + \sen(2\theta')\cos(\varphi') xz + \sen(2\theta')\sen(\varphi') yz \biggr)\Biggr) + \\&
    + \l(\frac{r_Q^2}{r'^4}\r)\Biggr(\frac{2\mu r}{\mu^2} - \biggr(\l(\cos^2(\varphi')\cos^2(\theta') + \sen^2(\varphi')\r)x^2 + (\sen^2(\varphi')\cos^2(\theta') + \\& \cos^2(\varphi'))y^2 + \sen^2(\theta')z^2 +  \l(\cos^2(\theta')\sen(2\varphi') - {\sen(2\varphi')}\r)xy + \\ & - \sen(2\theta')\cos(\varphi')xz - \sen(2\theta')\sen(\varphi')yz\biggr) \Biggr)\Biggr]\vast\} \qquad.
    \label{eq:potentialRNspherical}
\end{aligned}
\end{equation}}

We can also express the local system coordinates ${x,y,z}$ of the interacting system
in its own respective local spherical coordinates ${r,\theta,\varphi}$ and the expression is somewhat shortened and may be written as
{\footnotesize{
\begin{equation}
    \begin{aligned}
     V(\vec{r}) = V_0(r) \vast\{& 1 +
    \frac{1}{12}\Biggr\{\frac{r'^2}{r_Q^2 + r' (r' - r_s)} \biggr[\mu r \biggr(  \left( \cos(\theta  ) \cos(\theta) - \cos(\varphi' - \varphi) \sin(\theta') \sin(\theta) \right)^2  + \\& + \cos(\varphi' - \varphi) \sin(2\theta') \sin(2\theta) \biggr) - 1 \biggr] + \frac{r r_Q^2}{\mu r'^4}  \Biggr[ 2 - \mu r  \biggr( \cos^2(\theta) \sin^2(\theta')  - \cos(\theta) \cos(\varphi' - \varphi) \times  \\ & \times \sin(2\theta') \sin(\theta) + \sin^2(\theta) \left( \cos^2(\theta') \cos^2(\varphi' - \varphi)  + \sin^2(\varphi' - \varphi) \right) \biggr) \Biggr]\Biggr\}\vast\} \quad.
    \label{eq:potentialRNsphericalSPHERICAL}
\end{aligned}
\end{equation}
}}

\begin{figure}[!h]
    \centering
    \includegraphics[width=0.4\textwidth]{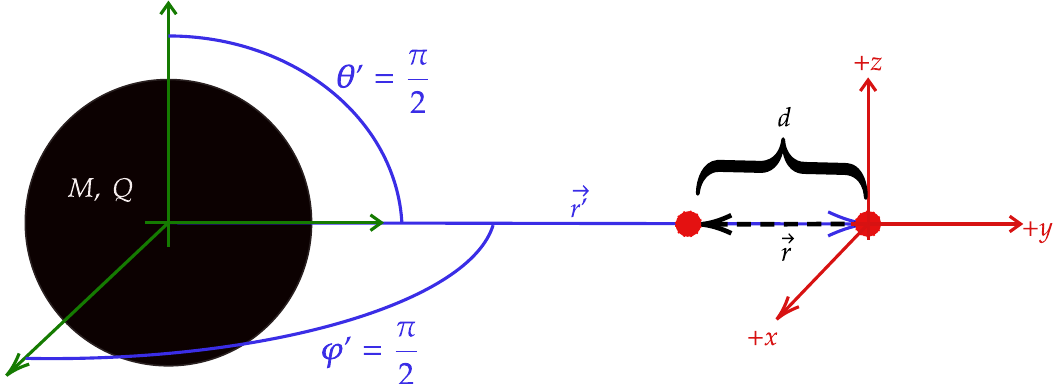}
    \caption{Representation of the relative coordinates of the interacting system with $\vr = (x=0,y=- d,z=0)$
    in relation to the charged black hole for the case of position $\theta' = \varphi' = \pi/2$.}
    \label{fig:casoI}
\end{figure}

\begin{figure}[!h]
    \centering
    \includegraphics[width=0.2\textwidth]{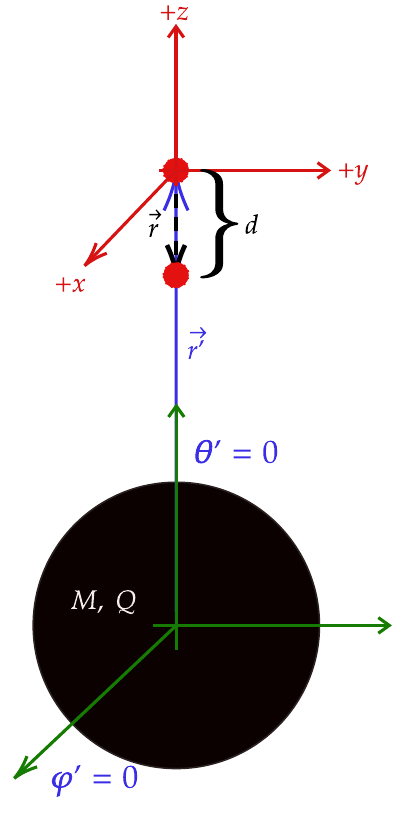}
    \caption{Representation of the relative coordinates of the interacting system with $\vr=(x=0,y=0,z= -d)$
    in relation to the charged black hole for the case of position $\theta' = \varphi' = 0$.}
    \label{fig:casoII}
\end{figure}

The RN black hole has a spherical symmetry, so it may seem sensible to think that the calculated corrections present the same kind of symmetry, that is, if changing from a relative position given by the coordinates $\{ \vec{r}^\prime,\theta', \varphi'\}$ to a rotated one determined by $\{ \vec{r}^\prime,  \theta'+\Delta\theta', \varphi' +\Delta\varphi' \}$, even with the alteration in the angular coordinates of the interacting system, the result is expected to be the same. In other words, the corrections determined by eq. \ref{eq:potentialRNspherical} and \ref{eq:potentialRNsphericalSPHERICAL} seem to depend on the angles $\theta'$ and $\varphi'$, but this behavior is only apparent, due to the mixing of the interacting and gravitational coordinate systems. Taking the examples represented in  \autoref{fig:casoI} and \autoref{fig:casoII} it becomes clear.

Considering the set of coordinates shown in \autoref{fig:casoI} 

\begin{equation}
    V(\vec{r}) = V_0(r) \vast\{1 + \frac{1}{12}\Biggr[
    \frac{r_Q^2}{r'^3 r_s-r'^4-r'^2 r_Q^2}\Biggr(\frac{\mu r}{\mu^2} - d^2\Biggr) + 
    \l(\frac{r_Q^2}{r'^4}\r)\biggr(\frac{2\mu r}{\mu^2} \biggr) \Biggr]\vast\}, \label{eq:casoIresult}
\end{equation}


\noindent
and for the arrangement shown in \autoref{fig:casoII} we have
\begin{equation}
    V(\vec{r}) = V_0(r) \vast\{ 1 + \frac{1}{12}\Biggr[
    \frac{r_Q^2}{r'^3 r_s-r'^4-r'^2 r_Q^2}\Biggr(\frac{\mu r}{\mu^2} - \biggr( z^2 \biggr)\Biggr) + \l(\frac{r_Q^2}{r'^4}\r)\Biggr(\frac{2\mu r}{\mu^2} \Biggr)\Biggr]\vast\} \   ,
    \label{eq:casoIIresult}
\end{equation}  
that gives the same result as eq. \ref{eq:casoIresult}.

So, if the interacting system is aligned with one axis of symmetry of the black hole with $y=d$ we have 
\begin{equation}
    V(\vec{r}) = V_0(r) \vast\{ 1 + \frac{1}{12}\Biggr[
    \frac{r_Q^2}{r'^3 r_s-r'^4-r'^2 r_Q^2}\Biggr(\frac{\mu r}{\mu^2} - d^2 \Biggr) + \l(\frac{r_Q^2}{r'^4}\r)\Biggr(\frac{2\mu r}{\mu^2} - \l(r^2-d^2\r) \Biggr)\Biggr]\vast\} \quad.
    \label{YukawaPotentialRNrd}
\end{equation}

\section{Results}


As far as the expression for $V(\vr)$ has been obtained, we can explore this result in order to determine  the effect of the structure of the spacetime by calculating the magnitude of the corrections.
So, in this section, we will study these corrections and how they may affect the potential, considering different sets of parameters.
We can therefore notice from eq.  \ref{YukawaPotentialRNrd} that these corrections depend on three parameters related to the gravitational source; two of them
from the charged black hole itself (charge $Q$ and mass $M$) and the distance to the interacting system $\l(r'\r)$, that appears encoded in two terms:
\begin{gather}
       R_{(1)} = \frac{r_Q^2}{r'^3 r_s-r'^4-r'^2 r_Q^2} = R_{rr}' \label{r1correction} \ , \\ 
     R_{(2)} = \frac{r_Q^2}{r'^4} = \frac{R'_{\theta\theta}}{r'^2} \label{r1r2corection} \quad .
\end{gather}

First, we bring our attention to possible boundaries for the charge parameter, whereas following the cosmic censorship hypothesis \cite{Penrose:1969pc} and gravitational collapse physical motivations \cite{carroll2003spacetime} there is a limit charge available for the black hole
\begin{equation}
    Q_{\mathrm{lim}} = \sqrt{4\pi G M^2} = \sqrt{4\pi} \frac{M}{m_P} \quad,
    \label{cargalimite}
\end{equation}
which is labeled as the extreme RN black hole and can be used to estimate how many elementary charges it consists of. In natural units this elementary charge is
$e = \sqrt{4\pi\a} \approx 0.30282$.

So, we can have an estimate of the number $n_{\mathrm{lim}}$ of elementary electric charges $e$ at the extreme case that is 
\begin{equation}
    n_{\mathrm{lim}} = \frac{Q_{\mathrm{lim}}}{e} = \frac{1}{\sqrt{\a}} \frac{M}{m_P} \quad,
\end{equation}
with $\a$ being the fine structure constant \footnote{$\a \approx 137^{-1}$ and $\frac{1}{\sqrt{\a}} \approx 11.7$, which leads to the conclusion that if the black hole mass has the minimum value of $1 m_P$ the correspondent charged black hole has a maximum limit electric charge around $11.7 e$}.

\begin{figure}[!h]
  \centering
  \begin{subfigure}{0.46\textwidth}
    \includegraphics[width=\textwidth]{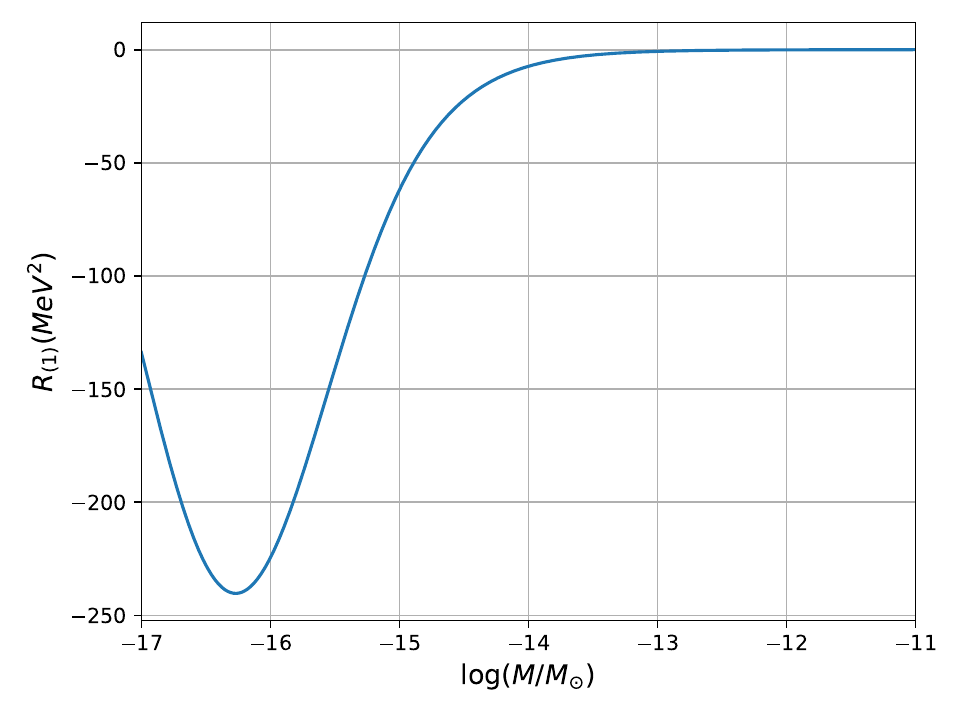}
    \caption{$R_{(1)}$ values in which $Q=0.99999Q_{\mathrm{lim}}$ at $r^{\prime}=r_+ + 10\mathrm{fm}$.}
    \label{fig:R1CorrectionMass}
  \end{subfigure}
  \hfill
  \begin{subfigure}{0.46\textwidth}
    \includegraphics[width=\textwidth]{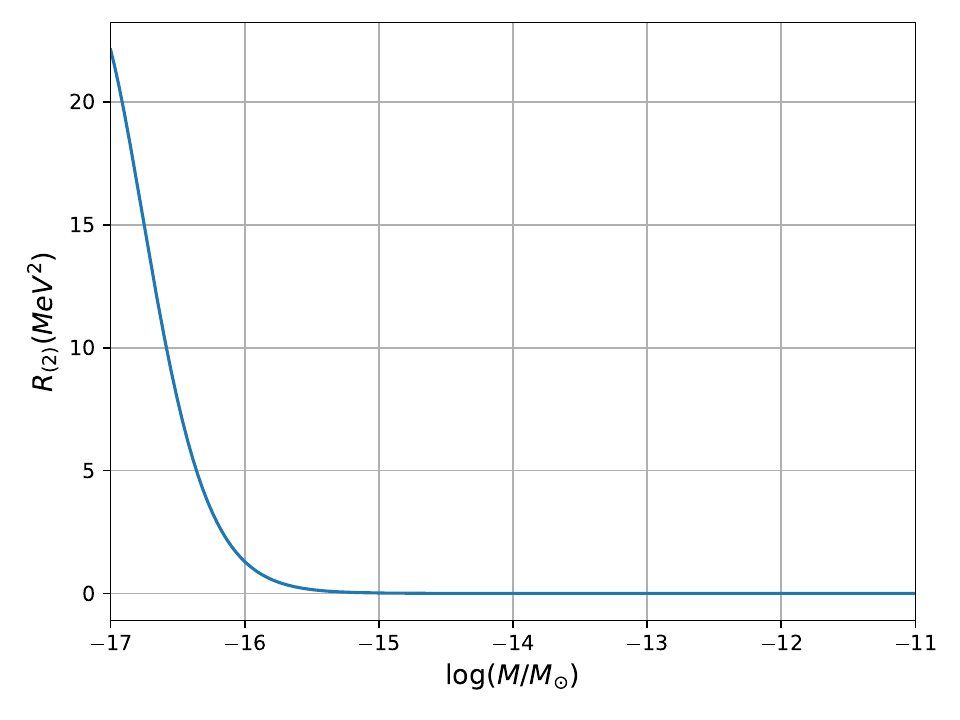}
    \caption{$R_{(2)}$ values in which $Q=0.99999Q_{\mathrm{lim}}$ at $r^{\prime}=r_+ + 10\mathrm{fm}$.}
    \label{fig:R2CorrectionMass}
  \end{subfigure}
    \caption{$R_{(1)}$ and $R_{(2)}$ correction terms for different mass values of RN black holes close to the external event horizon and near the limit charge.}
  \label{fig:CorrecoesMassas}
\end{figure}

Searching for the largest correction possible we can make an analysis for charges close to the extreme case and 
supposing the interacting system located
very near to the correspondent event horizon relative coordinate $r'_{Q_{\mathrm{lim}}} = r_s / 2$, since both terms coming from the Ricci tensor (eq. \ref{r1correction} and \ref{r1r2corection}), in absolute values, increase with the charge and also increase as $r'$ diminishes, as expected. We must also notice that $R_{(1)}$ diverges at the event horizon coordinates given by
\begin{equation}
    r_{\pm} = \frac{r_s}{2} \pm \sqrt{\l( \frac{r_s}{2} \r)^2 - r_Q^2} \quad,
\end{equation}
for any electric charge value, representing the limit of the formalism, as it is shown in \autoref{fig:correcoesrlinha}, for three solar masses (3$M_{\odot}$),
{ and it has different effects on the Yukawa potential depending on the location of \( r' \) relative to the event horizon radii \( r_{+} \) and \( r_{-} \). When \( r' \) is either greater than \( r_{+} \) or less than \( r_{-} \), \( R_{(1)} \) is negative, which implies that the Yukawa potential is weakened if \( \frac{r}{\mu} > d^2 \), and it is strengthened if \( \frac{r}{\mu} < d^2 \). In the region where \( r' \) is between \( r_{-} \) and \( r_{+} \), \( R_{(1)} \) is positive, which implies that the potential is strengthened if \( \frac{r}{\mu} > d^2 \), and it is weakened if \( \frac{r}{\mu} < d^2 \). Additionally, the second-order correction, \( R_{(2)} \), is always positive and non-divergent for all values of \( r' \), which implies that the potential is strengthened if \( \frac{2r}{\mu} > r^2 - d^2 \) and it is weakened if \( \frac{2r}{\mu} < r^2 - d^2 \).


The second-order correction \( R_{(2)} \) is consistently positive and non-divergent for all values of \( r' \), resulting in a strengthening of the potential if \( \frac{2r}{\mu} > r^2 - d^2 \) and a weakening if \( \frac{2r}{\mu} < r^2 - d^2 \).

\begin{figure}[!h]
  \centering
  \begin{subfigure}{0.46\textwidth}
    \includegraphics[width=\textwidth]{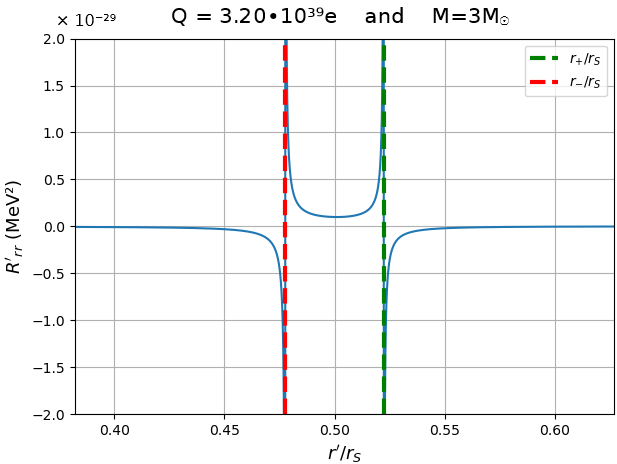}
    \label{fig:R1CorrectionMass}
  \end{subfigure}
  \hfill
  \begin{subfigure}{0.46\textwidth}
    \includegraphics[width=\textwidth]{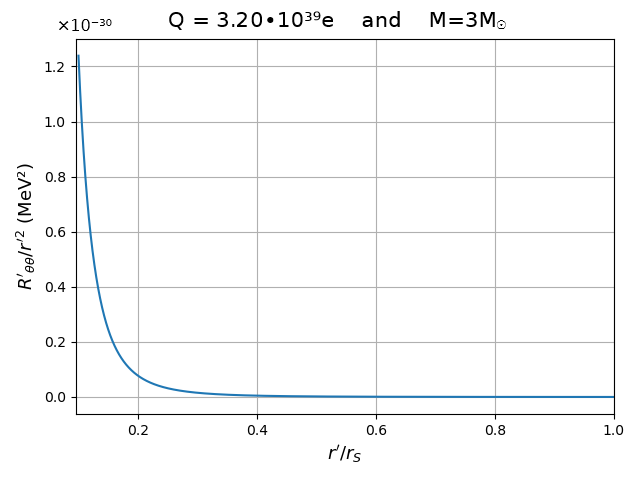}
    \label{fig:R2CorrectionMass}
  \end{subfigure}
    \caption{$R_{(1)}$ and $R_{(2)}$  for different radial distances with $M=3M_{\odot}$ and $Q=0.999Q_{\mathrm{lim}}$.}
  \label{fig:correcoesrlinha}
\end{figure}

We can vary the mass of the black hole in order to estimate the largest corrections available in an arbitrary vicinity of the event horizon. These results are shown in \autoref{fig:CorrecoesMassas}, where we can see considerable values for both correction terms ($\sim 1 \textrm{MeV}^2$) for $M \sim 10^{-15}M_{\odot}\sim 10^{16} \text{kg}$, which is, for comparison ends, equivalent to a hundredth of usual masses of asteroids in the solar system asteroid belt \cite{PitjevaAsteroid2018}, many orders smaller than observed and estimated masses of typical astrophysical black holes, demonstrating that these corrections are non-negligible for primordial or very low mass black holes \cite{HawkingLowMassBH:1971ei, CarrEarlyBigBangBlackHoles:1974nx,alonso2024}. As we can see from the numerical results, the first order spacetime curvature corrections are much smaller than 1 for most systems, eliminating the need of higher-order corrections in eq. \ref{PotentialCurvedSpaceInts}, which will be many orders of magnitude smaller.}

\begin{figure}[!h]
  \centering

  \begin{subfigure}{0.46\textwidth}
    \includegraphics[width=\textwidth]{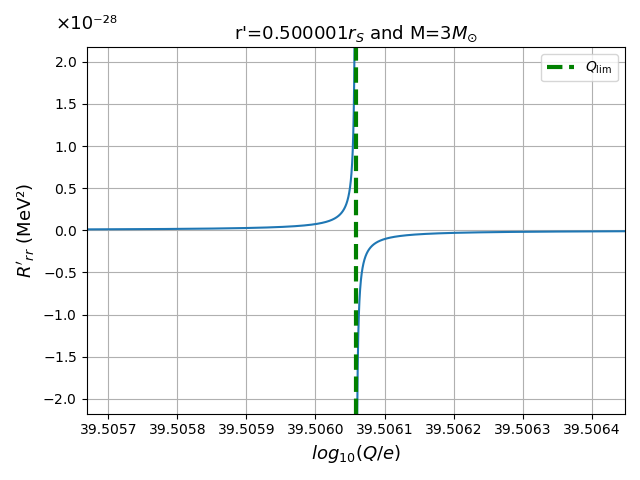}
    \caption{$R_{(1)}$ at $r' = 0.500001 r_{S}$.}
    \label{fig:R1CorrectionMass}
  \end{subfigure}
  \hfill
  \begin{subfigure}{0.46\textwidth}
    \includegraphics[width=\textwidth]{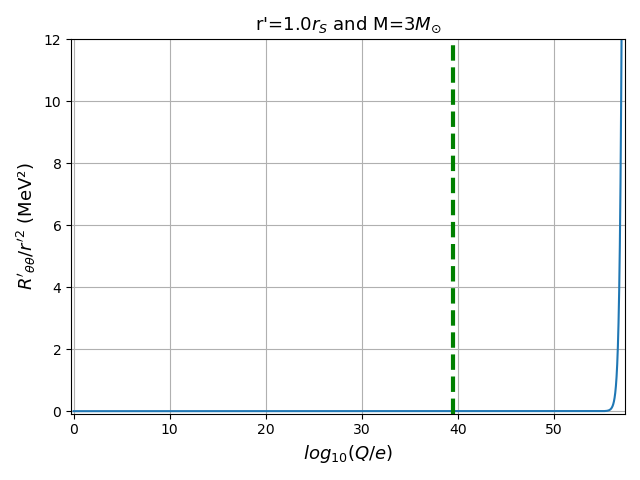}
    \caption{$R_{(2)}$ at $r' = r_{S}$.}
    \label{fig:R2CorrectionMass}
  \end{subfigure}
    \caption{$R_{(1)}$ and $R_{(2)}$ correction terms as functions of the electric charge for $3M_{\odot}$.}
  \label{fig:CorrecoesCargas}
\end{figure}

We can also examine the correction parameters by fixing the mass value and varying the electric charge. As an example, we take a black hole mass of $3 M_{\odot}$.  In \autoref{fig:CorrecoesCargas} we show the variation of both terms with the charge as it approaches the limit value of $\sim 3 \times 10^{39} e$. For charges larger than this theoretical barrier, $R_{(1)}$ changes sign and rapidly goes to zero, while $R_{(2)}$ continues to increase and provide large corrections around $Q \sim 10^{58}e$.

As for a better understanding of the numerical values of the correction we make the \autoref{table:deltaV} with values of $\abs{\frac{\Delta V}{V_0}} = \abs{\frac{V(g_{\mu\nu})-V_0}{V_0}}$, where $V_0$ is given by eq. \ref{V0} and $V(g_{\mu\nu})$ by eq. \ref{eq:potentialRNsphericalSPHERICAL}  , for different sets of masses and charges for the perpendicular case ($\theta = 0, \varphi=3\pi/2$). In \autoref{table:1} we
show the values for $\Delta V =V(g_{\mu\nu})-V_0$
 (in MeV) taking into account the same angular configuration, evaluated at the effective potential minimum, $r_{\mathrm{min}}$ \footnote{$r_{\mathrm{min}}$ being defined as the $r$ coordinate where $V_{\mathrm{eff}}$ has a minimum: $\d_rV_{\mathrm{eff}}|_{r_{\mathrm{min}}}=0$.}.
\begin{table}[h!]
\centering
\begin{tabular}{cccccc}
\toprule[1.5pt]
$Q$ \textbackslash $M$  & \(M_\mathrm{min}\) & \(100~ M_\mathrm{min}\) & \(\Msun\) & \(M_\mathrm{BH} = 10\Msun\) & \(M_\mathrm{SMBH} = 10^{10}\Msun\) \\
\midrule[1.5pt]
\(1e\) & $6.29 \times 10^{-49}$ & $3.48 \times 10^{-54}$ & $3.04 \times 10^{-101}$ & $1.85 \times 10^{-105}$ & $1.62 \times 10^{-58}$ \\
$0.1 ~ Q_{\mathrm{lim}}$ & $3.23 \times 10^{-7}$ & $1.79 \times 10^{-8}$ & $1.18 \times 10^{-24}$ & $3.75 \times 10^{-27}$ & $1.81 \times 10^{-44}$ \\
$0.5 ~ Q_{\mathrm{lim}}$ & $9.66 \times 10^{-6}$ & $5.88 \times 10^{-7}$ & $3.47 \times 10^{-23}$ & $1.21 \times 10^{-25}$ & $3.18 \times 10^{-43}$ \\
$0.9 ~ Q_{\mathrm{lim}}$ & $6.54 \times 10^{-5}$ & $6.34 \times 10^{-6}$ & $2.25 \times 10^{-22}$ & $1.39 \times 10^{-40}$ & $1.03 \times 10^{-42}$ \\
$0.99999 ~ Q_{\mathrm{lim}}$ & $2.49 \times 10^{-4}$ & $7.44 \times 10^{-4}$ & $5.55 \times 10^{-22}$ & $1.69 \times 10^{-24}$ & $5.09 \times 10^{-42}$ \\
\bottomrule[1.5pt]
\end{tabular}
\caption{$\abs{{\Delta V}/{V_0}}$ at \{$r= r_{\mathrm{min}}, \theta=0, \varphi=3\pi/2$\} for different combinations of masses and charges at the closest sensible distance of every respective case ($r'=r_{+}+10\mathrm{fm}$).}
\label{table:deltaV}
\end{table}

\begin{table}[h!]
\centering
\begin{tabular}{cccccc}
\toprule[1.5pt]
$Q$ \textbackslash $M$  & \(M_\mathrm{min}\) & \(100~ M_\mathrm{min}\) & \(\Msun\) & \(M_\mathrm{BH} = 10\Msun\) & \(M_\mathrm{SMBH} = 10^{10}\Msun\) \\
\midrule[1.5pt]
\(1e\) & $2.17 \times 10^{-47}$ & $1.20 \times 10^{-52}$ & $-1.05 \times 10^{-99}$ & $-6.40 \times 10^{-104}$ & $5.59 \times 10^{-57}$ \\
$0.1 ~ Q_{\mathrm{lim}}$ & $1.11 \times 10^{-5}$ & $6.19 \times 10^{-7}$ & $-4.09 \times 10^{-23}$ & $1.29 \times 10^{-25}$ & $6.25 \times 10^{-43}$ \\
$0.5 ~ Q_{\mathrm{lim}}$ & $3.33 \times 10^{-4}$ & $2.03 \times 10^{-5}$ & $1.20 \times 10^{-21}$ & $4.18 \times 10^{-24}$ & $1.10 \times 10^{-41}$ \\
$0.9 ~ Q_{\mathrm{lim}}$ & $2.26 \times 10^{-3}$ & $2.19 \times 10^{-4}$ & $-7.76 \times 10^{-21}$ & $-6.35 \times 10^{-39}$ & $-3.56 \times 10^{-41}$ \\
$0.99999 ~ Q_{\mathrm{lim}}$ & $8.59 \times 10^{-3}$ & $2.57 \times 10^{-2}$ & $-1.92 \times 10^{-20}$ & $5.85 \times 10^{-23}$ & $1.76 \times 10^{-40}$ \\
\bottomrule[1.5pt]
\end{tabular}
\caption{$\Delta V$ (in \textrm{MeV}) only at \{$r= r_{\mathrm{min}}, \theta=0, \varphi=3\pi/2$\} for different combinations of masses and charges at the closest sensible distance of every respective case ($r'=r_{+}+10\mathrm{fm}$).}
\label{table:1}
\end{table}

It is important to notice that in the tables the values are taken at the closest distances to the black holes
in order to obtain larger corrections.
As far as they increase for smaller values of $r'$  
 the way to get closer to the black hole
 and at the same time keeping the system outside the event horizon defined by $r_{+}$ is probing smaller and smaller masses.
 So, the closest distance possible between the strong interaction source and the black hole must be such that
in the scale probed (10 $\mathrm{fm}$) they do not overlap. This condition
 will imply the lowest mass possible black hole, that we call $M_{\mathrm{min}}$, whose horizon distance is associated with the limit charge, $r'_{\Qlim} = r_S/2$. Such mass is given by
\begin{equation}
\frac{r_S}{2} = \frac{ M_{\mathrm{min}}}{m_P^2} \gtrsim 0.05 \textrm{MeV}^{-1} \sim 10^{-14} \mathrm{m} \quad,
\end{equation}
and then $M_{\mathrm{min}}= 6.682 \times 10^{-18} \Msun=1.3287 \times 10^{13} \textrm{kg}$ that provides the largest corrections  for $r'_{+}+10\ \textrm{fm}$. This fact may be observed in \autoref{table:deltaV} and \autoref{table:1}. For typical black holes
with masses $M_{\rm{BH}}$ or supermassive black holes with $M_\mathrm{SMBH}$,
the corrections are small and increase for black holes with small masses and large values of $Q$.

\begin{figure}[!h]
    \centering
    \includegraphics[scale=0.62]{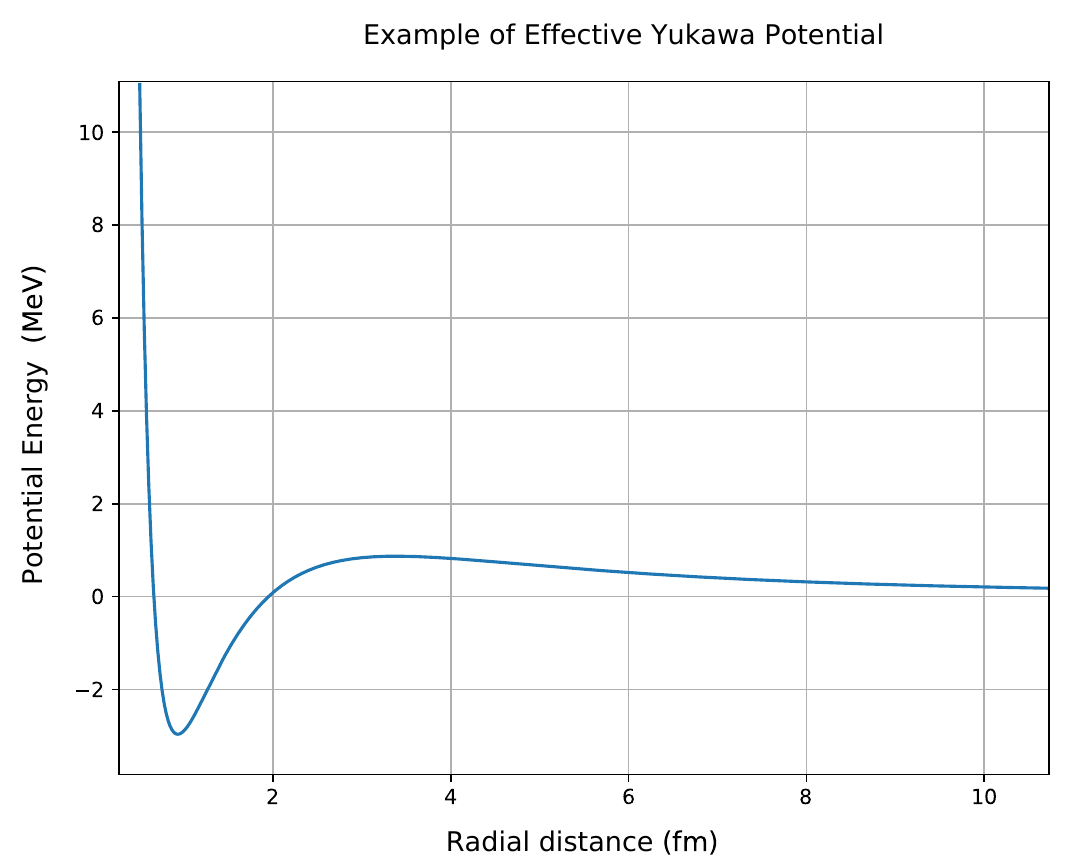}
    \caption{Effective potential energy for nucleon-nucleon interactions from eq. \ref{Veff} in the Minkowski space.}
    \label{fig:yukawaeffectiveplot}
\end{figure}

These results may also provide insights about how nuclear systems are affected by curved spaces. Examining \autoref{table:deltaV} and \autoref{table:1} it is possible to know how the parameters of the theory may affect the interaction. Observing that $V_0$ obtained with the $\phi^3$ theory has the same functional form of the central part of the Yukawa potential determined by a one-pion exchange in the nucleon-nucleon interaction, and that this kind of potential is widely used in non-relativistic calculations with the Schrödinger equation \cite{BookNNInteractionbrown}, it is possible to propose  
a corrected form of the effective central potential
\begin{equation}
    V_{\mathrm{eff}} = V(g_{\mu\nu}, \vr) + \frac{L^2}{2mr^2} \quad,
    \label{Veff}
\end{equation}
with $V(g_{\mu\nu}, \vr)$ given by eq. \ref{cartesianCurvedSpacePotentialResult}
which presents a potential well which characterizes bound states of nuclear interactions. For a comparison with the nuclear system we take the proton mass $m=m_p=938.272\mathrm{MeV}$, the pion mass
$\mu = m_\pi = 134.977\mathrm{MeV}$, $\lambda=4450\mathrm{MeV}$ and $L^2=2\hbar^2$, which in the Minkowski space is shown in \autoref{fig:yukawaeffectiveplot}.

\begin{figure}[!h]
    \centering
    \includegraphics{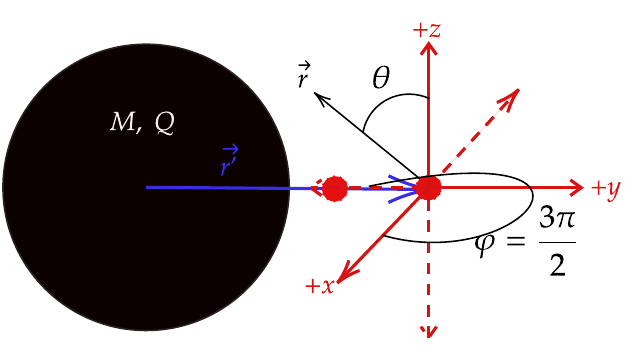}
    \caption{Pictorial representation for the position of an interacting system located on an axis in the $y \times z$ plane that means $\varphi=3\pi/2)$.}
    \label{fig:anglesplitrep}
\end{figure}

\newpage
First it is interesting to analyze how the effective potential behaves in terms of the relative angular coordinates. So, placing the particles in the $y \times z$ plan, with $y$ being the axis that connects both systems centers, we can simply plot $V(r, \theta, \varphi=3\pi/2)$ for different values of $\theta$ as shown in \autoref{fig:anglesplitrep}. From the quadratic dependence on the coordinates in the corrections of the potential, it manifests a symmetry at the quadrants of such plan, so we can just vary $\theta$ between $0$ and $\pi/2$ and see what the effect look like in this different directions. 

For this we take different sets of parameters relatively close to the exterior event horizon, in particular, at a  distance $r' \gtrsim r_{+} + 10{\text{fm}}$, such that the effects are considerable.

\begin{figure}[!h]
    \centering
    \includegraphics[scale=0.7]{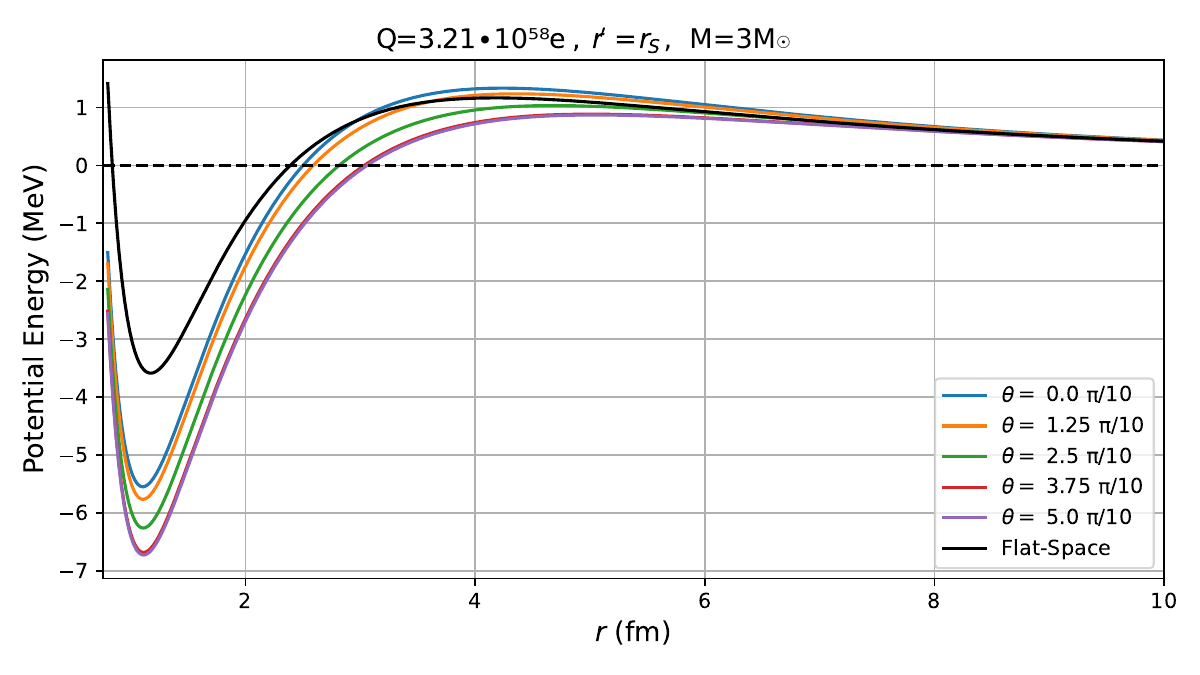}
    \caption{Effective potential as function of the radial distance $r$ for different values of $\theta$ in the case of $M=3M_{\odot}$ and  $Q \approx 10^{58} e \gg Q_{\mathrm{lim}}$ at $r'=r_S\approx8.86\mathrm{km}$.}
    \label{fig:AnglesSplitPotentialsLargeQ}
\end{figure}

    At the cases regarding stellar black hole masses, the smallest masses (and therefore largest contributions at the event horizon) are around $M = 3 M_{\odot}$, and will have non-negligible contributions only with electric charge outside the physical regime, for example, $Q \sim 10^{58}e$, far beyond $Q_{\mathrm{lim}}$, such that only the contribution from $R_{(2)}$ is non-negligible. At $r'=r_S$, we have \autoref{fig:AnglesSplitPotentialsLargeQ}, where we can see that, indeed, larger gravitational corrections as the angle $\theta$ goes to $\pi/2$, or, the relative position is closer to the line connecting the gravitational and strong interaction centers.
    
\begin{figure}[!h]
    \centering
    \includegraphics[scale=0.64]{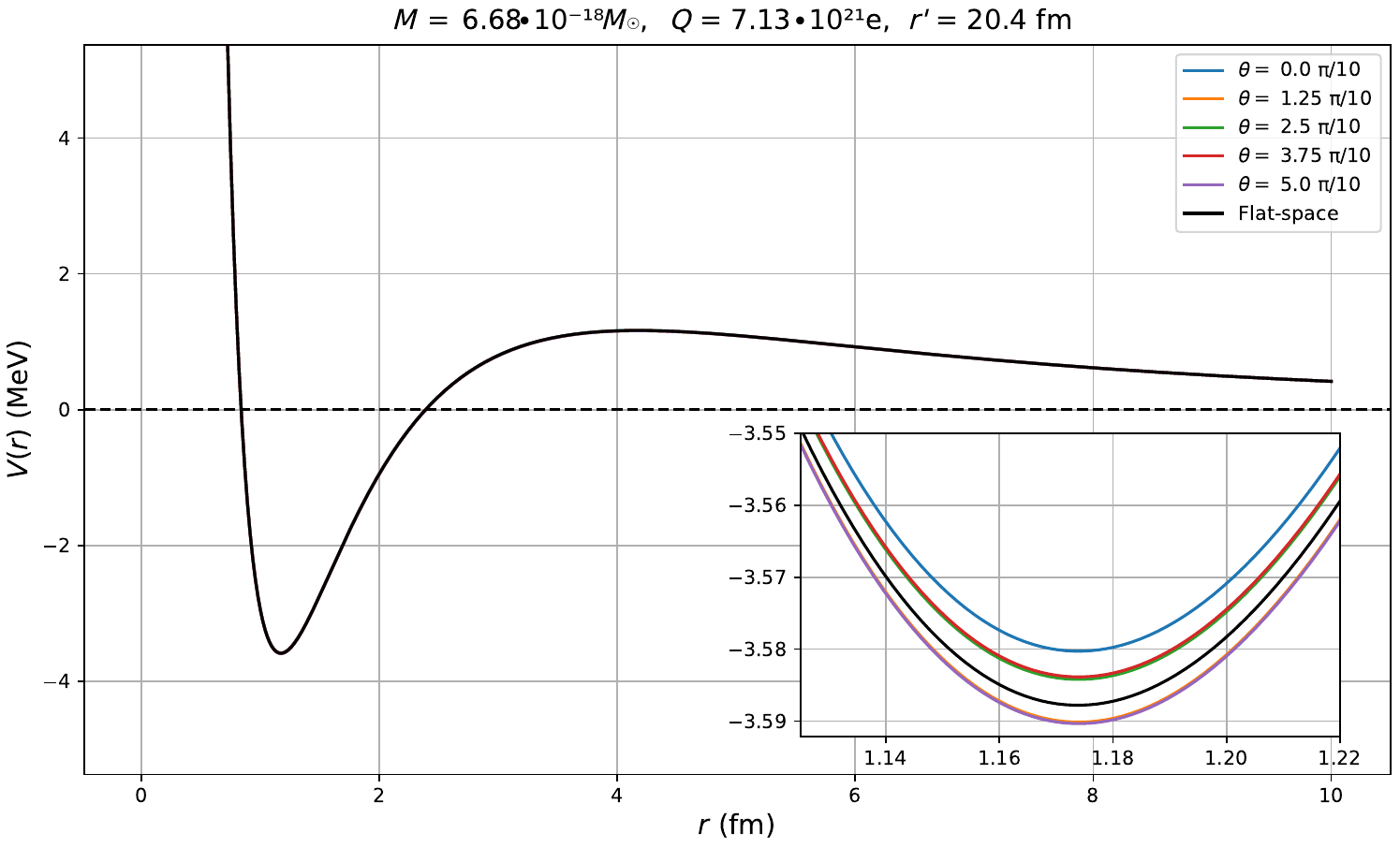}
    \caption{Effective potential as function of the radial distance $r$ for different values of $\theta$ in the case of $M=6.682 \times 10^{-18}\Msun \sim 10^{20}m_P$ and  $Q=0.999~Q_{\mathrm{lim}} \sim 10^{21}e$ at $r'=r_{+}+10~ \mathrm{fm} \sim 20~\mathrm{fm}$.}
    \label{fig:AnglesSplitPotentialsSmallMlimiar}
\end{figure}

The lowest mass scenario ($M_{\rm{min}}$) is shown in \autoref{fig:AnglesSplitPotentialsSmallMlimiar}, where we can see the potential being split whereas the difference for all relative angles $\theta$ are smaller than $0.01 \textrm{MeV}$ in the case of a charge around $10^{22}e$, equivalent to $99.9\%$ of $Q_{\mathrm{lim}}$, at the distance around $20 \textrm{fm}$. From the plot we can see that for $\theta=0$, where the two-particle system orientation is perpendicular to the black hole radial direction, the potential well is diminished, which represents a weakening of the interaction. In the $\theta=\pi/2$ case, when the two-particle system orientation is parallel to the black hole radial direction, the potential well is enlarged, which represents a strengthening in the interaction.

As we can see the spacetime parameters might affect the effective potential as a form of diminishing the potential well, maybe even eliminating it depending on their values. This means that if the system is located in a critic parameter volume in the parameters space \footnote{For instance, in the RN metric, it would imply in a $\{M,r',Q\}$ space.} the potential well and bound states may be broken. The critic parameters represent a surface this critic volume, that is, in the critic limit values of these parameters, which can have many possible configurations, the potential may have a saddle point, satisfying:
\begin{equation}
    \begin{aligned}
    \frac{d}{dr}V_{eff}(g_{\mu\nu},\theta=0,r,\varphi) = 0 \\
    \frac{d^2}{dr^2}V_{eff}(g_{\mu\nu},\theta=0,r,\varphi) = 0
    \end{aligned} \quad .
    \label{eq:saddlepointconditions}
\end{equation}
~
This kind of analysis and the determination of the regions in parameter space that present or not a potential well is left for future works.

A deeper analysis of the system regarding
its phenomenological aspects such as 
bound states and particle emission amplitudes may be obtained in a straightforward way
by solving the correspondent Schrödinger equation with the presented modified potential.


\section{Conclusions}

In this work we analyzed how a curved spacetime might affect a Yukawa-like
potential of an interacting quantum system generated by a spin-0 field exchange. 
In deriving this altered interaction potential, the propagator, when defined in an arbitrary metric other than Minkowski's, has corrections to its flat space counterpart expression, which in the Bunch-Parker local momentum representation \cite{BunchParkerProp, BunchProp2RenormPhi4:1981tr} is given by a series of terms that depend on the geometric quantities derived from the metric, such as the Ricci scalar and the components of the Ricci tensor evaluated at a localized spacetime region, and we must emphasize that in this framework corrections of vertices and external lines does not appear. This interesting result demonstrates that the curved space propagator encodes not only the 4-momentum and mass information of the particles, but also information about the geometry of the local spacetime manifold. It is important to emphasize that in this work we elaborated a method to study nuclear forces in general curved spaces and showed an example. The natural continuation of this investigation is to look for systems where the corrections for the potential become relevant, and the fundamental aspect of considering other theoretical approaches which are not based on the Riemann Normal Coordinates in order to verify the presence of other kinds of corrections. 

As a first test, we used a simple model that reproduces a Yukawa potential, the $\phi^3$ theory, which determines a potential for the interacting bosons. Applying the propagator result to the first Born approximation and the dominant-order scattering diagram, we extracted the corrected Yukawa potential up to the third order in curved space. From this result, it is possible to analyze different metrics of interest. In order to study these results, we considered the simplest black hole metric with a non-null Ricci tensor, the charged black hole, described by the Reissner-Nordström  metric.

Qualitatively, the effect of curvature is not trivial, since the radial symmetry is broken. Due to the non-symmetric curved background at the point of expansion, we have a different energy configuration available for different configurations of the quantum system position. In particular, we took interest in analyzing the scenarios of perpendicular and parallel to the line connecting the strong source to the black hole center. 

The formulation in the RN metric brings a dependence on the relative distance $r'$ from the quantum system to the center of the black hole and also on its mass $M$ and charge $Q$. Varying these three parameters we noticed that the curvature corrections are extremely small for astrophysical masses near the correspondent event horizon, even in the case near the extreme solution in which $Q=Q_{\mathrm{lim}}$, where the ratio between the gravitational correction, at the effective potential minimum, and the correspondent flat space potential value is on the order of $10^{-22}$ for a mass of $M_{\odot}$ with a charge of 0.99999 $Q_{\mathrm{lim}}$, and on the order of $10^{-4}$ for a mass of $M_\mathrm{min}$ with a charge of 0.99999 $Q_{\mathrm{lim}}$ as well.
We also noticed that these corrections may be observable when the mass of the black hole is small, such as the primordial black holes \cite{HawkingLowMassBH:1971ei}, \cite{CarrEarlyBigBangBlackHoles:1974nx} and the ones recently proposed in \cite{alonso2024}. As we discussed in the text, we can expect that corrections of the same magnitude appear when studying low energy nucleon-nucleon interactions. For the central part of the potential the corrections should be the same. 

This kind of corrections, even if are very small, affect any astrophysical system where the strong interactions are important. A typical example where these corrections could be observed is in the formulation of equations of state for nuclear matter within the framework of relativistic star models and the exploration of its possible implications using multi-messenger observational data \cite{ETL-Maselli:2020uol, ETL-Sabatucci:2023hpa,ETL-Tiwari:2023tkj,ETL-Rose:2023uui,ETL-Ozel:2016oaf,ETL-hamel:2008ca,ETL-Lattimer:2000nx}. Another possibility of observing this kind of effect is with the next-generation gravitational-wave detectors, which probably will offer refined constraints on nucleon interactions. These observations could indirectly probe spacetime-curvature-induced modifications to nuclear potentials, providing many elements to verify the effects proposed in this work.

Concurrently, Einstein's theory of general relativity and the exploration of quantum effects represent forefronts of modern theoretical research. The intersection of these domains remains a fertile ground for inquiry, with the reconciliation of quantum mechanics and gravity standing as a central challenge in theoretical physics.
Indeed, the incorporation of curved spacetime effects into the framework of the quantum field theory represents a frontier of contemporary research, with implications for our understanding of the physical theory at a deeper level.

\section{Acknowledgments}

We thank the Coordenação de Aperfeiçoamento de Pessoal de Nível Superior (CAPES), process number 88887.655373/2021-00, and Conselho Nacional de Desenvolvimento Científico e Tecnológico (CNPq) for the financial support.

\bibliographystyle{ieeetr}
\bibliography{main.bib}

\end{document}